\documentclass[prl,twocolumn,amsmath,amssymb,nofootinbib]{revtex4}
\usepackage{graphicx} 
\usepackage{amsmath}
\usepackage{amssymb}
\usepackage{bbold}
\usepackage{color}
\usepackage{hyperref}
\usepackage{amsthm}

\def\bea{\begin{eqnarray}}
\def\eea{\end{eqnarray}}
\def\<{\langle}
\def\>{\rangle}

\def\uada{{\uparrow\downarrow}}

\def\be{\begin{equation}}
\def\ee{\end{equation}}

\def\ss{{\text{s}}}

\begin{document}

\title{ A Quantum Computational Determination of  the Weak Mixing Angle\\ in the Standard Model}

\author{ Qiaofeng Liu$^{\, a}$, Ian Low$^{\, a,b}$, Zhewei Yin$^{\, a}$}
\affiliation{\vspace{0.1cm}
\mbox{$^a$ Department of Physics and Astronomy, Northwestern University, Evanston, IL 60208, USA}\\ 
\mbox{$^b$ High Energy Physics Division, Argonne National Laboratory, Lemont, IL 60439, USA}\\
 \vspace{-0.3cm}}

\begin{abstract}
The weak mixing angle $s_W$ is a fundamental constant in the Standard Model (SM) and measured at the $Z$ boson mass to be $\widehat{s}^2_W(m_Z) = 0.23129 \pm 0.00004$ in the $\overline{\rm MS}$ renormalization scheme, where  $m_Z=91.2$ GeV. On the other hand, non-stabilizerness -- the magic -- characterizes the computational advantage of a quantum system over classical computers. We consider the production of magic from  stabilizer initial states, which carry zero magic, in the 2-to-2 scattering of charged leptons in the SM at the tree level, which is mediated by the photon and the $Z$ boson.  Using the second order stabilizer R\'enyi entropy, and averaging over all 60 initial stabilizer states and the scattering  angle, we  compute and minimize the magic production as a function of $s^2_W$  in the M\o ller  scattering $e^-e^-\to e^-e^-$, which  is free of kinematic thresholds. At the centre-of-mass energy $\sqrt{s}=m_Z$, there is a unique minimum in  magic production at  $\mathbf{s}^{2}_W(m_Z)=0.2317$, which agrees with the measured  $\widehat{s}^2_W(m_Z)$  at the sub-percent level.  At higher energies, the magic-minimizing  $\mathbf{s}^{2}_W$  continues to agree with the empirical value at the percent level or better, up to 10 TeV. The finding suggests the electroweak sector of the SM tends to generate  minimal  quantum resources from the computational viewpoint.
\end{abstract}

\maketitle

\section{Introduction} 

What separates quantum from classical?  Schr\"odinger proclaimed in 1935 that entanglement is the most prominent feature of quantum mechanics \cite{Schrodinger1935}. However, from the computational point of view, entanglement is not the main source for quantum speedup, as demonstrated by  the Gottesman-Knill theorem \cite{gottesman1998heisenbergrepresentationquantumcomputers,Gottesman:1999tea,Aaronson:2004xuh}, which states that certain quantum circuits involving maximally entangled states  can be simulated efficiently, in polynomial time, using classical algorithms. Bravyi and Kitaev introduced the concept of non-stabilizerness, or the magic, to characterize  the computational advantage in a quantum system as the second layer of quantumness \cite{Bravyi:2004isx}. Moreover, magic states are  essential  for universal quantum computation.

Recently   increasing attention has been directed toward understanding  fundamental interactions from the perspective of quantum information. Examples include the appearance of enhanced symmetries \cite{Beane:2018oxh,Low:2021ufv,Liu:2022grf,Carena:2023vjc,Liu:2023bnr,Hu:2024hex,Chang:2024wrx,Kowalska:2024kbs,McGinnis:2025brt,Busoni:2025dns,Carena:2025wyh,Hu:2025lua}, a novel area law in particle scatterings \cite{Aoude:2024xpx,Low:2024mrk,Low:2024hvn}, as well as the production of magic states in nuclear and particle physics  \cite{Robin:2024bdz,Chernyshev:2024pqy,White:2024nuc,Liu:2025frx,Aoude:2025jzc,Liu:2025qfl,Gargalionis:2025iqs}.  There are also efforts to estimate parameters in the SM from the entanglement perspective. Ref.~\cite{Cervera-Lierta:2017tdt} obtained  $s^2_W=1/4$ by requiring maximal entanglement in $e^+e^-\to \mu^+\mu^-$ at certain scattering angles, while Ref.~\cite{Thaler:2024anb} considered deriving the pattern of flavor mixing in the SM from entanglement minimization. However, from the computational viewpoint, entanglement alone is not sufficient to guarantee quantum resources.

In this work we would like to initiate a study on the following question: Is the SM of particle physics special from the perspective of quantum computation? Broadly speaking, given that quantum computing  holds great potential to process information and simulate  systems which are too complex for classical computers, it is important to study how and where to generate the quantum resource for quantum computation. Moreover, is quantum complexity an inherent property of fundamental interactions  or an emergent phenomenon?

\begin{figure}
    \centering
    \includegraphics[width=1\linewidth]{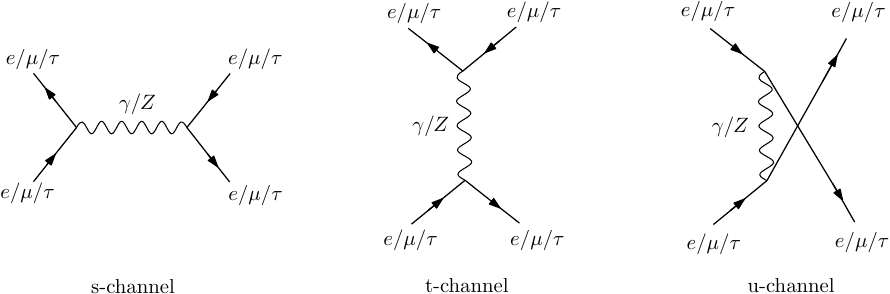}
    \caption{The $s/t/u$-channel Feynman diagrams for charged lepton scatterings in the SM. }
    \label{fig:feyn}
\end{figure}

 In this work we study the magic production in 2-to-2 scattering of charged leptons, $\ell^\pm=\{e^\pm, \mu^\pm, \tau^\pm\}$, at the tree-level in the SM, which include both the photon and the $Z$ boson exchanges.  Since the $Z$  coupling to $\ell^\pm$ depends on the weak mixing angle $s^2_W$, we pay particular attention to the magic production as a function of $s^2_W$. The scattering process in general involves three classes of Feynman diagrams, the $s$/$t$/$u$-channels as shown in {Fig.~\ref{fig:feyn}}. However, the $Z$ boson exchange in the $s$-channel   introduces a kinematic threshold due to resonant $Z$ productions. Therefore the magic production in processes involving the $s$-channel, such as $\ell^+\ell^-\to \ell^+\ell^-$, is somewhat unstable near threshold. Consequently, we will focus on the M\o ller scattering $\ell^-\ell^-\to \ell^-\ell^-$, which does not contain the $s$-channel and is free of  threshold effects. 
 
 We proceed to minimize the magic production with respect to $s^2_W$ in $\ell^-\ell^-\to \ell^-\ell^-$ and  find a unique minimum at  $\mathbf{s}^2_W$  that agrees impressively well with  the experimental value  in the $\overline{\rm MS}$ scheme. At  $\sqrt{s}=m_Z$ the agreement is at the sub-percent level, while at higher energies the agreement continues at the percent level or better, after including the $\sqrt{s}$ dependence in the magic production and the renormalization group running of gauge couplings. The reason we focus on energies at around $m_Z$ or higher is because the sensitivity of the scattering amplitudes on the weak mixing angle is maximal in this energy regime. At energies much lower than $m_Z$, the photon exchange becomes dominant and the magic production reverts to that in Quantum Electrodynamics, which was studied in Ref.~\cite{Liu:2025qfl}.

\section{Stabilizer R\'enyi Entropy}

Characeterizing and quantifying the magic  has become an important topic  in quantum  computation \cite{Mari:2012ypq,Veitch_2012,Emerson:2013zse,Bravyi_2019,PhysRevLett.123.020401,PRXQuantum.3.020333,Warmuz:2024cft,Leone:2021rzd}.  Here we provide a lightning review. More  detailed discussions can be found in  Refs.~\cite{Nielsen:2012yss,Leone:2021rzd}.  Among the several measures of magic, the stabilizer R\'enyi entropy (SRE) \cite{Leone:2021rzd} has gained popularity   due to its ease to implement and being measurable in experiments \cite{Oliviero:2022bqm,Haug:2023ffp}. The SRE was originally proposed for qubits and  has since been generalized to qudits \cite{Wang:2023uog}. 

Let $\mathcal{P}_n$ be the set of $n$-qubit Pauli strings,
\begin{equation}
    \mathcal{P}_n = \left\{P_1 \otimes P_2\otimes\dots\otimes P_n \right\} , \quad  P_i \in 
    \left\{ I, X, Y, Z \right\} \ ,
\end{equation}
where $I=\sigma^0$, $X=\sigma^1$, $Y=\sigma^2$ and $Z=\sigma^3$. Given a pure quantum state $|\psi\rangle$, the SRE of   order $\alpha$  is defined   as
\bea
    M_\alpha \left(\left|\psi\right\rangle\right) 
    &=& \frac{1}{1-\alpha} \log\, \sum_{P\in \mathcal{P}_n} \frac1d  \left\langle\psi\vert P \vert \psi \right\rangle^{2\alpha} \ ,
\eea
where $d = 2^n$ is the dimensionality of the $n$-qubit Hilbert space. For a stabilizer state there exist $d$ mutually commuting Pauli strings for which $\langle\psi|P| \psi \>^2=\pm 1$ and vanishes otherwise.  The SRE is defined such that $M_\alpha=0$ for a stabilizer state.  For a product state the SRE is additive: $M_\alpha \left(\left|\psi\right\rangle\otimes|\phi\>\right) = M_\alpha \left(\left|\psi\right\rangle\right)+M_\alpha \left(|\phi\>\right)$. For  a non-stabilizer state, $M_\alpha> 0$. In the following, we focus on two-qubit systems, $d=4$, and choose $\alpha=2$,
\bea
M_2 (| \psi \>) =- \log\, \sum_{P\in \mathcal{P}_2} \frac14  \left\langle\psi\vert P \vert \psi \right\rangle^{4} \ .
\eea
In  quantum resource theory the SRE with $\alpha\ge 2$ has also been shown to be a good quantitative measure of the quantum resource \cite{Veitch_2014,RevModPhys.91.025001,Leone:2024lfr,Bittel:2025yhq}. Moreover,  the maximum of $M_2$ for two-qubit states is  shown to be \cite{Liu:2025frx}
\be
M_2 \le \log\frac{16}{7}\approx 0.827\ . 
\ee
Basic forces in nature do not seem to generate states with the maximal magic  easily in 2-to-2 scatterings \cite{Liu:2025qfl,Gargalionis:2025iqs}. By contrast, maximally entangled states are produced abundantly in high energy scattering processes.

An important property of the SRE is that it is invariant under Clifford operations: $|\psi\>\to C|\psi\>$, where $C\in\{{\rm Hadamard}, {\rm Phase}, {\rm CNOT}$\}. This is intimately connected to the Gottesman-Knill theorem \cite{gottesman1998heisenbergrepresentationquantumcomputers,Gottesman:1999tea,Aaronson:2004xuh}  which states that a quantum computation can be simulated efficiently in the computational basis using classical algorithms if  the operation and measurement involve only  the Clifford  gates. Under these conditions some highly entangled quantum states would not provide quantum speedup  over classical algorithms. In addition,  Clifford gates alone are not sufficient for universal quantum computation, which requires  a non-Clifford T-gate or, equivalently, magic states \cite{Bravyi:2004isx}.

\section{The Weak Mixing Angle} 

The electroweak sector of the SM is based on the gauge symmetry $SU(2)_L\times U(1)_Y$ with the gauge coupling strengths $g$ and $g^\prime$, respectively. Upon spontaneous symmetry breaking, only the $U(1)_{\text{em}}$ subgroup remains unbroken, which describes the electromagnetic force. The corresponding massless gauge boson is the  photon $A_\mu$. Then the $SU(2)_L\times U(1)_Y$ gauge couplings $\{g, g^\prime\}$ can be rewritten as  the weak mixing angle $s_W$ and the positron electric charge $e$:
\bea
\label{eq:swmsbar}
s_W &=& \frac{g^\prime}{\sqrt{g^2+g^{\prime\,2}}}\ , \\
e &=& \frac{g g^\prime}{\sqrt{g^2+g^{\prime\,2}}} = g s_W \ , 
\eea
The weak mixing angle also describes the rotation of the neutral gauge bosons from the gauge eigenstate $(W^3, B)$ to the mass eigenstate $(Z, A)$:
\be
\left(\begin{array}{c} Z\\
                             A \end{array}\right) = \left(\begin{array}{cc} c_W & -s_W \\
                                                                                                  s_W & c_W \end{array}\right)
                                                                                                  \left(\begin{array}{c} W^3\\
                             B \end{array}\right)\ .
\ee
The photon coupling to the charged fermion is
\be
{\cal L}_{A\bar{f}f}= -e\, Q_f \, \bar{f}\gamma^\mu f \, A_\mu\ ,
\ee
where $Q_f$ is the electric charge of the fermion in unit of $e$.  In the broken sector the massive gauge bosons are $W_\mu^\pm$ and $Z_\mu$, with masses $m_W={80.4}$ GeV and $m_Z=91.2$ GeV. The $Z$ coupling to the fermion is
\be
{\cal L}_{Z\bar{f}f} = -\frac{e}{2 s_W c_W} \, \bar{f}\gamma^\mu (g_V^\ell -g_A^\ell \gamma^5) f\, Z_\mu \ ,
\ee
where 
\be
\label{eq:gvga}
g_V^\ell = T_3^f - 2 Q_f \, s_W^2 \ , \qquad g_A^\ell = T_3^f \ .
\ee
For charged leptons $T_3^\ell = -1/2$ and $Q_\ell=-1$.

Beyond the leading order in perturbation, quantum effects introduce a logarithmic scale dependence in the gauge coupling constants through the procedure of renormalization. In the modified minimal subtraction ($\overline{\rm MS}$) scheme one simply promotes Eq.~(\ref{eq:swmsbar}) to include the scale dependence of the gauge couplings in the $\overline{\rm MS}$ scheme  \cite{ParticleDataGroup:2024cfk}:
\be
 s_W^2(\mu) = \frac{g^{\prime\,2}(\mu)}{g^2(\mu)+g^{\prime\,2}(\mu)} \ .
 \ee
 For  electroweak processes the scale $\mu$ is often chosen to be $m_Z$. The world average from various measurements is  \cite{ParticleDataGroup:2024cfk}
 \be
 \label{eq:smsw}
 \widehat{s}_W^2(m_Z)=0.23129 \pm 0.00004 \ .
 \ee
 Notice that when $s_W^2=1/4$, the $Z$ coupling to the charged leptons becomes purely axial as can be seen from Eq.~(\ref{eq:gvga}). This is the reason that entanglement is maximized at $s^2_W=1/4$ in $e^+e^-\to \mu^+\mu^-$ scattering \cite{Cervera-Lierta:2017tdt}.

\section{Magic in M\o ller Scattering}

The M\o ller scattering in the SM, $\ell^-(p_1)\ell^-(p_2)\to \ell^-(k_1)\ell^-(k_2)$, is mediated by the photon and the $Z$ boson in the $t$- and $u$-channels shown in Fig.~\ref{fig:feyn}. The fermions  are  qubits in the spin space and we choose the incoming beam axis in the centre-of-mass (CM) frame to project the spins, for {\em both} the incoming and the outgoing leptons. Specifically, this ``Lab basis''  is defined by
\bea
\hat{z} = \hat{p}_1, \qquad \hat{y} = \frac{\vec{k}_1 \times \vec{p}_1}{|\vec{k}_1 \times \vec{p}_1|}, \qquad \hat{x} =  \hat{y} \times \hat{z}.
\eea
We will adopt the computational basis to describe the spin orientations of the two particles: $\{|00\>,|01\>,|10\>,|11\>\}$ where, for instance, the 2-qubit state $|01\>\equiv |\uada\>$ means that the spin of particle 1 is in the $+z$-direction, while the spin of particle 2 is in the $-z$-direction. This  choice of coordinate system is very common in high energy collider physics \cite{Gainer:2011xz} and was adopted in calculating the magic production in QED \cite{Liu:2025qfl}. 

An alternative choice is the helicity basis  projecting the spin along the direction of motion. The coordinate rotation from the Lab basis to the helicity basis is not a Clifford operation and, therefore, will modify the magic. This is similar to the fact that a $SU(4)$ operation in the computational basis that is not a local, single-qubit operation will  change the amount of entanglement. The choice of Lab basis is in line with the observation that on a quantum computer, the computational basis is usually the norm, and it generally requires certain non-Clifford gates to move away from the computational basis. Moreover, the Gottesmann-Knill theorem is originally phrased in the computational basis.

The spin state of the incoming particles is chosen to be one of the 60 stabilizer states $|\psi_\ss\>_{i}$, $i = 1,2, \cdots, 60$, for two-qubit systems. The full list can be found in Ref.~\cite{Liu:2025qfl}. To highlight that entanglement alone is not sufficient to characterize computational resources, we point out that the first 36 stabilizer states listed in Ref.~\cite{Liu:2025qfl} are not entangled while the rest are maximally entangled. On the other hand, all 60 stabilizer states have zero magic.

The scattering amplitude in M\o ller scattering  can be calculated using standard techniques in quantum field theory. At a particular scattering angle $\theta$,  the spin configuration of the outgoing leptons is given by the amplitude and represented by a state vector $|\tilde{\psi}\>$ in the computational basis. Since this is considered a projective measurement onto a fixed angle,  we use the normalized state $|\psi\> = |\tilde{\psi}\>/(\<\tilde{\psi}|\tilde{\psi}\>)^{1/2}$ to compute the magic produced at the angle $\theta$. In the Lab basis, the Dirac spinors for the fermions do not carry any azimuthal $\phi$ dependence; neither do the resulting amplitudes. 

Next is to compute the SRE averaged over all 60 initial stabilizer states \cite{Robin:2024bdz} and the final state solid angle,
\bea
{\cal M}_2(\theta) &\equiv& \frac{1}{60} \sum_{i=1}^{60}   M_2(|\psi_\ss\>_{i},\theta)\ , \\
\<{\cal M}_2\> &\equiv& \frac1{4\pi} \int  {\cal M}_2(\theta) \, \sin\theta\, d\theta\,d\phi\ .
\eea
$\<{\cal M}_2\>$ represents the ability of SM to generate computational resources in the M\o ller scattering from stabilizer initial states. In the SM, it only depends on the lepton mass $m_\ell$, the $Z$ boson mass $m_Z$ and width $\Gamma_Z$, and the weak mixing angle $s_W$; the dependence on the electric charge drops out when using the normalized state $|\psi\>$ for the outgoing particles. Since the CM energy we are considering, at $m_Z$ or higher, is much larger than the lepton masses, $\sqrt{s}\gg m_\ell$, $\ell=e, \mu,\tau$, the magic production has almost no dependence on whether we are scattering $e$, $\mu$, or $\tau$. In the end, at a fixed CM energy, $\<{\cal M}_2\>$  is only sensitive to the weak mixing angle $s_W$.

\begin{figure}[t]
    \centering
  \includegraphics[width=0.8\linewidth]{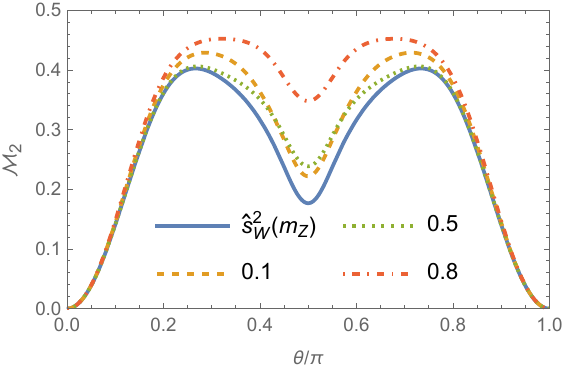}
    \caption{Angular distribution of magic production ${\cal M}_2(\theta)$ for M\o ller scattering $e^- e^-\to e^-e^-$ at $\sqrt{s}=m_Z$ with $s_W^2=\widehat{s}_W^2(m_Z), 0.1, 0.5,$ and 0.8. }
    \label{fig:m2vsth}
\end{figure}  

\begin{figure*}[th]
    \centering
    \includegraphics[width=0.85\linewidth]{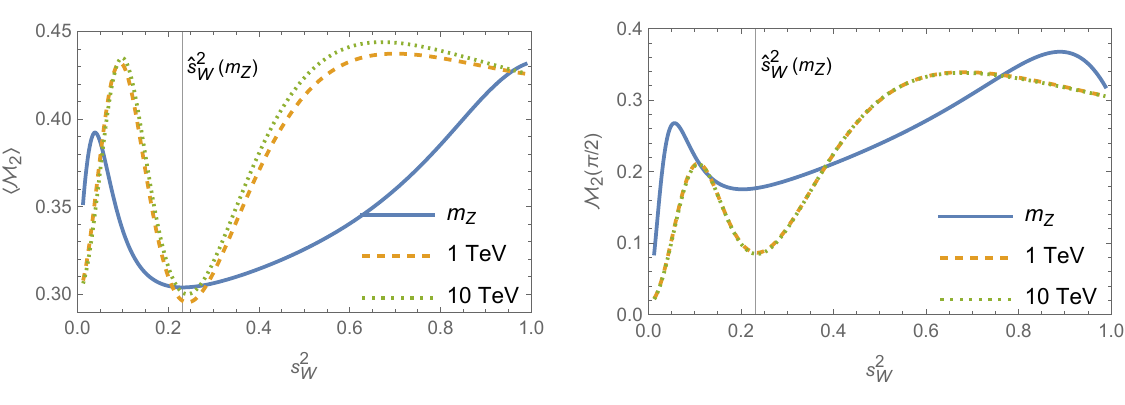}
    \caption{ Magic production in M\o ller scattering $e^-e^-\to e^-e^-$ as a function of the weak mixing angle $s_W^2$ for CM energies $\sqrt{s}=m_Z$, 1 TeV and 10 TeV.  }
    \label{fig:swdep}
\end{figure*}  

In Fig.~\ref{fig:m2vsth} we show ${\cal M}_2(\theta)$ at $\sqrt{s}=m_Z$ for several values of $s_W^2$. Two features are prominent: 1) the  magic production as a function of $\theta$, averaged over stabilizer initial states, has a minimum at $\theta=\pi/2$ for different values of $s_W^2$ and
2) the overall magic production seems to increase when $s_W^2$ deviates from the $\widehat{s}_W^2(m_Z)$, the SM value. Then in Fig.~\ref{fig:swdep} we plot both $\<{\cal M}_2\>$ and ${\cal M}_2(\pi/2)$ as a function of  $s_W^2$, for $\sqrt{s}=m_Z$, 1 TeV and 10 TeV. Interestingly, in both cases,  the magic production has a minimum at $s_W^2$ that is very close to the SM value $\widehat{s}_W^2(m_Z)$.

\begin{figure}[t]
    \centering
  \includegraphics[width=0.8\linewidth]{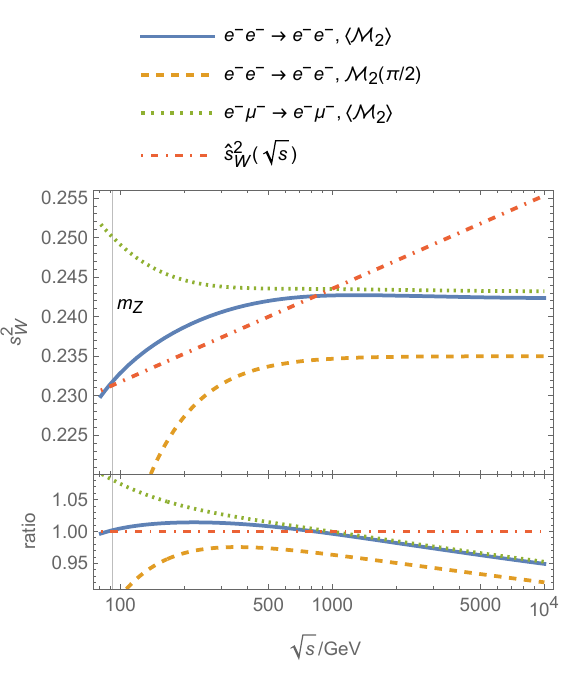}
    \caption{The magic minimizing $s_W^2$ for both ${\cal M}_2(\pi/2)$ and $\<{\cal M}_2\>$ as a function of $\sqrt{s}$, as compared to the RGE of the empirical SM value. The $e^-e^-\to e^-e^-$ process contains both $t$- and $u$-channels, while $e^-\mu^-\to e^-\mu^-$ has only $t$-channel.}
    \label{fig:m2vsroots}
\end{figure}

At first glance one might suspect that  the location of the minimum is exactly at $s_W^2=1/4$, in which case the $Z$ coupling to the charged lepton is purely axial and the final state entanglement is maximal at certain scattering angles when the initial state is unentangled, as discovered in  Ref.~\cite{Cervera-Lierta:2017tdt}. In this regard, we compute the precise location of the minimum in the magic production and compare with the SM value, taking into account both the $\sqrt{s}$ dependence in the scattering amplitude of M\o ller scattering as well as the renormalization group evolution (RGE) of the gauge couplings, which gives rise to the energy dependence in $\widehat{s}_W^2(\sqrt{s})$ \cite{Erler:2017knj,ParticleDataGroup:2024cfk}. The result is shown in Fig.~\ref{fig:m2vsroots}, where we present the location of the minimum in both $\<{\cal M}_2\>$ and ${\cal M}_2(\pi/2)$ as a function of $\sqrt{s}$. For comparison we also include the energy dependence of the SM value $\widehat{s}_W^2(\sqrt{s})$. Moreover, since the $e^-e^-\to e^-e^-$ process contains both the $t$- and the $u$-channel, 
we further compute the magic production in the process $e^-\mu^-\to e^-\mu^-$, which contains only the $t$-channel contribution.

It is clear from Fig.~\ref{fig:m2vsroots} that for M\o ller scattering the magic-minimizing $s_W^2$ never reaches 1/4 over the range of $\sqrt{s}$ we considered. At $\sqrt{s}=m_Z$, the minimum of $\<{\cal M}_2\>$ is located at
\be
\label{eq:sw2min}
\mathbf{s}_W^2(m_Z) = 0.2317 \ .
\ee
Comparing with the SM value at $m_Z$ in Eq.~(\ref{eq:smsw}), the agreement is at the sub-percent level, which does not appear to be due to the $Z$ boson coupling being purely axial to the fermion at $s_W^2=1/4$. In addition, the minimum of ${\cal M}_2(\pi/2)$, the magic production at $\theta=\pi/2$, is further away from the empirical value when comparing with the minimum of $\<{\cal M}_2\>$,  although the agreement with $\widehat{s}_W^2(m_Z)$ is still quite well. At higher energies, $\mathbf{s}_W^2(\mu)$ continues to agree well with $\widehat{s}_W^2(\mu)$ up to $\mu=10$ TeV, and they never deviate from each other by more than a few percent. In fact, at energies between $m_Z$ and $\mu\sim 1$ TeV, the two values are consistent to a very high precision. In Fig.~\ref{fig:m2vsroots} we stop at the $W$ boson mass at $m_W=80.4$ GeV because the RGE of $\widehat{s}_W^2$ changes drastically due to the adoption of an effective field theory where the $W$ and $Z$ bosons are integrated out \cite{ParticleDataGroup:2024cfk}, making the comparison with the magic minimizing value much less direct.

The minimum of magic production from the $t$-channel contribution, as captured by the $e^-\mu^-\to e^-\mu^-$ process, is still in the vicinity of the M\o ller scattering. One can also plot the magic minimizing $s_W^2$ for $e^-\mu^-\to \mu^-e^-$, which only contains the $u$-channel contribution, and the result is exactly the same as the $e^-\mu^-\to e^-\mu^-$ case. This is not surprising, as the two processes differ in interchanging the final state particles, which amounts to a SWAP gate acting on the two qubits in the final state. This is a Clifford operation that does not change magic. The notably different behavior of $e^-e^-\to e^-e^-$ from $e^-\mu^-\to e^-\mu^-$ near $\sqrt{s} = m_Z$ in Fig.~\ref{fig:m2vsroots} thus indicates that the superposition of the $t$- and $u$-channel contributions in M\o ller scattering is essential for the sub-percent agreement at $\sqrt{s} = m_Z$.

In the calculations we retain the full effects of lepton masses. The value in Eq.~(\ref{eq:sw2min}) is for both the electron and the muon M\o ller scattering. For the $\tau$ lepton, the value for $\mathbf{s}_W^2$ become 0.2316.

\section{Discussions}

In this work we presented a quantum-computational determination of the weak mixing angle in the SM at energies near or above the weak scale. Focusing on the M\o ller scattering $e^-e^-\to e^-e^-$, which is free of kinematic threshold at $\sqrt{s}=m_Z$, we minimize the magic production in the final state with respect to $s_W^2$, when the initial states are stabilizer  states with zero magic. At $\sqrt{s}=m_Z$, the minimum of the magic production, averaged over all 60 stabilizer initial states and the scattering angle $\theta$, turns out to be consistent with the measured value at $m_Z$ at the sub-percent level. If only considering the minimum of magic production at $\theta=\pi/2$, the corresponding agreement is not
as good, but still within the order of 10\%. After taking into account the $\sqrt{s}$ dependence of the scattering amplitude, as well as the renormalization group evolution of the weak mixing angle, the magic-minimizing $s_W^2$ continues to be consistent with the SM value within a few percent, up to $\sqrt{s}=10$ TeV.

There is a long history of efforts in constructing  physics-beyond-the-SM (BSM) theories to predict values of $s_W^2$ close to the empirical value.
These models involve a combination of grand unification, additional symmetry and interactions beyond the $SU(3)\times SU(2)_L\times U(1)_Y$ in the SM, the presence of new particles, or even extra spacetime dimensions \cite{Weinberg:1971nd,Georgi:1974sy,Pati:1974yy,Dimopoulos:1981zb,Mohapatra:1983aa,Dimopoulos:2002mv,Chacko:2002vh}. These BSM theories predict $s_W^2=3/8$ or $1/4$ at an energy scale (sometimes much) higher than the weak scale and rely on the RGE to bring $s_W^2$ close to the empirical value at $m_Z$. The present approach is very distinct from all the previous efforts in that it adopts the quantum computational point of view  and minimizes the SRE, which is a measure of the quantum advantage over a classical computer. No hypothetical interactions and particles are introduced.

The observation that the empirical value of the weak mixing angle at  $m_Z$ sits very close to a value which minimizes the production of non-stabilizerness in 2-to-2 scattering of charged leptons from stabilizer initial states indicates that the electroweak sector of SM likes to produce minimal quantum advantage from the computational perspective, which seems to be consistent with the findings in Refs.~\cite{Liu:2025qfl,Gargalionis:2025iqs} that the fundamental forces in nature are not efficient in generating quantum resources necessary for quantum speedup. The finding opens up a new frontier to understand the SM  from the perspective of quantum resources in quantum computation. Since we only considered tree-level diagrams in this work, it would also be interesting to study how higher order corrections in perturbation would affect these findings.

\section{Acknowledgments}
We thank Yingying Li, Fabio Maltoni, Jesse Thaler, Sokratis Trifinopoulos, Carlos Wagner and Martin White for comments on the manuscript.  This work is supported in part by the U.S. Department of Energy under contracts DE-AC02-06CH11357 (Argonne), DE-SC0023522 (Northwestern), DE-SC0010143 (Northwestern) and
No. 89243024CSC000002 (QuantISED Program).

\bibliography{ref}

\end{document}